\documentclass[pra,twocolumn,showpacs,superscriptaddress]{revtex4-1}
\usepackage{epsfig,graphicx,amssymb,color,bm}
\usepackage{amsmath,float}
\usepackage[colorlinks, linkcolor=blue, urlcolor=blue, citecolor=blue, unicode=true]{hyperref}
\usepackage{setspace}
\usepackage{url}
\newcommand{\eref}[1]{Eq.~(\ref{#1})}
\newcommand{\esref}[1]{Eqs.~(\ref{#1})}
\newcommand{\fref}[1]{Fig.~\ref{#1}}
\begin{document}

\title{Quantum correlation enhanced weak field detection in optomechanical system}
\author{Wen-Zhao Zhang}\thanks{zhangwz@csrc.ac.cn}
\affiliation{Beijing Computational Science Research Center (CSRC), Beijing 100193, China}
\author{Li-Bo Chen}
\affiliation{School of Science, Qingdao Technological University, Qingdao 266033, China}
\author{Jiong Cheng}
\affiliation{Department of Physics, Ningbo University, Ningbo 315211, China}
\author{Yun-Feng Jiang }
\affiliation{Department of Mechanical and Aerospace Engineering, University of California San Diego, 9500 Gilman Drive, La Jolla, CA 92093, USA}

\begin{abstract}
We propose a theoretical scheme to enhance the signal-to-noise ratio in ultrasensitive detection with the help of quantum correlation.
By introducing the auxiliary oscillator and treated as an added probe for weak field detection, the additional noise can be greatly suppressed and the measurement accuracy may even break the standard quantum limit.
We use the magnetic field as an example to exhibit the detection capability of our scheme.
The result show that, comparing with the traditional detection protocol, our scheme can have higher signal-to-noise ratio and better detection accuracy.
Furthermore, the signal intensity detection curve shows a good linearity.
Our results provide a promising platform for reducing the additional noise by utilizing quantum correlation in ultrasensitive detection.
\end{abstract}

\maketitle
\section{introduction}
Quantum technology has a very important application prospect in ultrasensitive detection, such as mass sensor \cite{apl.106.121905}, force sensor \cite{NJP.19.083022} and quantum gyroscopes \cite{PhysRevA.95.012326}, etc.
The use of quantum correlation can provide substantial enhancements for detecting and imaging the weak signals in the presence of high levels of noise and loss \cite{PhysRevLett.114.080503,science.321.1463,PhysRevA.92.043816,NP.14.160,sr.6.31095,PhysRevLett.115.243603}.
Emission entangled photon to improve the detection signal-to-noise ratio (SNR) in quantum illumination \cite{PhysRevLett.114.080503,science.321.1463},
via optical parametric amplification to enhance the quality of ghost imaging  \cite{PhysRevA.92.043816},
intracavity squeezing to enhance the precision of a position measurement \cite{PhysRevLett.115.243603}.
These protocols show the superiority of using quantum correlation in sensitive detection.

Optomechanical systems have been widely used in weak field sensing due to the direct relationship of the mechanical displacement and the field intensity \cite{PS.2.259}.
More importantly, as a good probe, mechanical oscillators may couple with various kinds of field, such as electric field \cite{nature.464.697},  magnetic field \cite{PhysRevLett.108.120801}, and even gravitational waves \cite{PhysRevLett.113.151102}.
Therefore, optomechanics can be used to detect these signals and it is an important candidate for ultrasensitive detection
Such systems exploit the huge susceptibility around the resonance frequency of oscillators with excellent mechanical quality factor $Q_m$, combined with high-sensitivity interferometric measurements \cite{PhysRevA.97.033833,QO}.
Due to the competitive relation between photon shot noise and quantum back-action noise, the detector based on photomechanical system has a standard quantum limit (SQL), which limits the further improvement of measurement accuracy.
It has been shown that, optomechanical systems can be utilized as detectors that has noise performance beyond the SQL by suppressing the additional noise with the help of quantum coherence.
Such noise suppression is achieved with the quantum effects associated with the optomechanical systems, including squeezing \cite{PhysRevLett.115.211104,PhysRevLett.113.151102}, entanglement \cite{Anp.385.757,PhysRevLett.114.080503}, optical high-order correlation \cite{PhysRevLett.120.020503} and so on.

Up to now, most of the measurement schemes are focus on breaking SQL \cite{RevModPhys.86.1391}.
However, the use of quantum correlation of the optical cavity to cancellation the additional noise will usually reduce the amplification of the signal \cite{PhysRevA.92.043817}.
The performance of a quantum detector based on optomechanical system should not only be evaluated by the suppression of its additional noise, but also should take into account the signal amplification as well as the linearity of the detector response.
For the reasons given above, in this paper, an optomechanical dual-probe port scheme (OMDP) is proposed.
By using the quantum correlation between input ports, we can greatly reduce quantum noise and surpass the SQL in the field measurement.
In addition, the scheme can exhibit high SNR and good ability to resist environment temperature.
The paper is organized as follows.
In Sec. II, we first introduce the model, then give the quantum Langevin equations and calculate the spectrum of fluctuations of the output light, after that, we give the analytic expression for the weak field sensitivity.
In Sec. III, we study the additional noise and SNR in the weak field detection.
In Sec. IV, we take the magnetic field as an example to show the superior capability of our scheme.
In Sec. V, we summarize our main conclusions.

\section{Model and Hamiltonian}
\begin{figure}
  \centering
  \includegraphics[width=7cm]{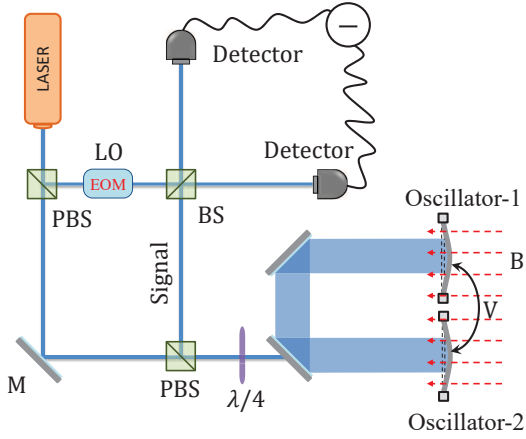}\\
  \caption{The schematic diagram of the weak field detection system.
  The monochromatic cavity field is coupled with two movable mirrors, and two coupled movable mirrors are used as probes to induce weak field with strength $B$.
  The laser is split into a signal beam and a local oscillator (LO).
  The signal beam is used to drive the cavity.
  The LO is phase-modulated with an electro-optical modulator (EOM) and used to perform homodyne detection with the output signal beam. }\label{fig1}
\end{figure}
As shown in \fref{fig1}, the weak field detector is composed of a composite optomechanical system and a homodyne detection system.
The Hamiltonian of composite optomechanical system can be described as
\begin{eqnarray}
  H_S &=& \omega_c a^{\dag}a +\sum_{j=1,2}[\frac{\omega_j}{2}(p_j^2+q_j^2)-g_jq_j a^{\dag}a]+Vq_1q_2,\nonumber\\
\end{eqnarray}
where $a$ is the bosonic operators for the optical mode with frequency $\omega_c$.
$p_j$ and $q_j$ are the position and momentum operator of the j-th mechanical mode with frequency $\omega_{mj}$.
The single-photon coupling coefficient of the optomechanical interaction is $g=(\omega _{c}/L)\sqrt{\hbar /(2m\omega _{m})}$.
The standard continuous-wave driving can be described as $H_d =\varepsilon[a \exp(-i\omega_d t)+a^{\dag} \exp(i\omega_d t)]$, where $\omega_d$ is the angular frequency of the laser and $\epsilon$ is the cavity driving strength, given by $\epsilon \equiv 2\sqrt{P \kappa_{ex}/(\hbar \omega_d)}$, with $P$ being the input power of the laser and $\kappa_{ex}$ being the input rate of the cavity.
$V$ is the strength of two mechanical coupling and can be realized via a substrate-mediated interaction \cite{PhysRevLett.115.017202}.
The strength of the detection field is $B$, which could be electric field, magnetic field \cite{PhysRevLett.107.207210} and gravitational field \cite{PhysRevLett.113.151102} (An example will be discussion in Sec IV).
Through the design, we can make the mechanical oscillators produce small displacement $q_j$ in the weak field.
The corresponding mechanical response coefficient of the oscillators in the weak field is $\xi_j$.
The Hamiltonian of the mechanical oscillators surround by the classical field can be written as $H_b = -\sum_{j=1,2}B\xi_j q_j$.

Under strong driving condition, we can linearize the equations of motion around the steady state with $a \rightarrow \langle a \rangle +\delta a$, $p_j\rightarrow \langle p_j \rangle +\delta p_j$, $q_j\rightarrow \langle q_j \rangle +\delta q_j$.
In the rotating frame under input laser frequency $\omega_d$, the quantum Langevin equations can be obtained,

\begin{eqnarray}
 \delta \dot{a} &=& -(i\Delta'+\frac{\kappa}{2})\delta a+\sum_{j=1,2}i G_j \delta q_j+\sqrt{\kappa}a_{in},\\
  \delta \ddot{q_j} &=&-\omega_{mj}^2\delta q_j+\omega_{mj}(G_j \delta a^{\dag}+G_j^*\delta a)\nonumber\\
  &&-\omega_{mj}V\delta q_{3-j}-\omega_{mj}\gamma_j \delta p_j+\omega_{mj}F_{in,j},
\end{eqnarray}
where $\Delta'=\Delta-\sum_{j=1,2}g_j\langle q_j\rangle$ denotes the driving modified detuning of the cavity, $G_j=g_j\langle a\rangle$ denotes the linearized coupling.
$F_{in,j}=\sqrt{\gamma_j}P_{th,j}+\xi_j B$ is the input term of the mechanical oscillator, $F_{th,j}$ is the noise operator.
Introducing the Fourier transformation, the operator dynamic of the system in frequency domain becomes,
\begin{eqnarray}
 \delta a(\omega) &=& \chi_c[i G_1 \delta q_1(\omega)+i G_2 \delta q_2(\omega)+\sqrt{\kappa}a_{in}(\omega)]\\
  \delta q_j(\omega)&=&\chi_{mj}[G_j \delta a^{\dag}(-\omega)+G_j^*\delta a(\omega)\nonumber\\&&-V\delta q_{3-j}(\omega)+F_{in,j}(\omega)]
\end{eqnarray}
where $\chi_c=[i(\Delta'-\omega)+\kappa/2]^{-1}$,  $\chi_{mj}=[\omega_{mj}-\omega^2/\omega_{mj}-i\gamma_j\omega/\omega_{mj}]^{-1}$ are susceptibilities
of cavity and mechanical oscillators, respectively.
For the slow varying field or the steady field, the input term can be written as
$F_{in,j}(\omega)=\sqrt{\gamma_j}P_{th,j}(\omega)+\delta(\omega) \xi_j B$.
Using the standard input-output relation $O_{out}=\sqrt{\kappa }O-O_{in}$ and considering the hommodyne measurement shown in Fig.~\ref{fig1}, we have the output operator
\begin{eqnarray}
M_{out}(\omega ) &=&i[a_{out}^{\dag }(-\omega )e^{-i\theta }-a_{out}(\omega)e^{i\theta }]  \label{eqout} \\
&=&A(\omega )a_{in}(\omega )+B(\omega )a_{in}^{\dag }(-\omega )\nonumber\\&&+C(\omega)F_{in,1}(\omega)+D(\omega)F_{in,2}(\omega),  \nonumber
\end{eqnarray}
where, $\theta$ is an controllable phase which can be used to reduce the additional noise and has been studied in Ref. \cite{PhysRevA.90.043848}.
For simplicity, we choose $\theta=0$, the coefficients in \eref{eqout} can be written as
\begin{eqnarray}\label{eqps}
A(\omega ) &=& i(1-\kappa \chi_c)+\nonumber\\&&\frac{i\kappa \chi_c (|G|^2\chi_c+G^{*2}\chi_c^{\dag})(2V\chi_{m1}\chi_{m2}-\chi_{m1}-\chi_{m2})}{D_e}, \nonumber\\
B(\omega ) &=& -i(1-\kappa \chi_c^{\dag})+\nonumber\\&&\frac{i\kappa \chi_c^{\dag} (|G|^2\chi_c^{\dag}+G^{*2}\chi_c)(2V\chi_{m1}\chi_{m2}-\chi_{m1}-\chi_{m2})}{D_e}, \nonumber \\
C(\omega ) &=&\frac{i\sqrt{\kappa}(G \chi_c+G^* \chi_c^{\dag})(V\chi_{m1}\chi_{m2}-\chi_{m1})}{D_e},  \nonumber\\
D(\omega ) &=&\frac{i\sqrt{\kappa}(G \chi_c+G^* \chi_c^{\dag})(V\chi_{m1}\chi_{m2}-\chi_{m2})}{D_e}.
\end{eqnarray}
the denominator $D_e=i(V^2 \chi_{m1}\chi_{m2}-1)+|G|^2(\chi_c-\chi_c^{\dag})(2V\chi_{m1}\chi_{m2}-\chi_{m1}-\chi_{m2})$, the cavity susceptibility satisfy the relationship $\chi_c^{\dag}=\chi_c^*(-\omega)$.

\section{Weak field detection}
Assumption that the two mechanical oscillators have the same response rate $\xi_1=\xi_2=\xi$.
\eref{eqout} can be rewritten as
\begin{eqnarray}\label{eqout1}
M_{out}(\omega ) &=&A(\omega )a_{in}(\omega )+B(\omega )a_{in}^{\dag }(-\omega )+C(\omega)\sqrt{\gamma_1}P_{th,1}(\omega)\nonumber\\
&&+D(\omega)\sqrt{\gamma_2}P_{th,2}(\omega)+[C(\omega)+D(\omega)]\delta(\omega)\xi B,
\end{eqnarray}
According to the above equation, the coefficient of the weak field intensity ($B$) directly represents the amplification characteristic of the output signal to the detection signal.
Therefore, the amplification coefficient of the system can be defined as $A_p=|C(\omega)+D(\omega)|$.
Analogously, the other items without $B$ in the output operator can be regarded as noise.
To obtain the relationship between the output signal and the noise, we can rewrite \eref{eqout1} as
\begin{widetext}
\begin{eqnarray}
\frac{M_{out}}{C(\omega)+D(\omega)}&=&\frac{A(\omega)}{C(\omega)+D(\omega)}a_{in}(\omega)+\frac{B(\omega)}{C(\omega)+D(\omega)}a_{in}^{\dag}(\omega)
+\frac{C(\omega)\sqrt{\gamma_1} P_{th,1}+D(\omega)\sqrt{\gamma_2} P_{th,2}}{C(\omega)+D(\omega)}+\delta(\omega)\xi B, \label{mout1}
\end{eqnarray}
\end{widetext}
thus the additional noise can be defined as
\begin{eqnarray}
F_{add}&=&\frac{A(\omega)}{C(\omega)+D(\omega)}a_{in}(\omega)
+\frac{B(\omega)}{C(\omega)+D(\omega)}a_{in}^{\dag}(\omega)\nonumber\\&&
+\frac{C(\omega)\sqrt{\gamma_1} P_{th,1}+D(\omega)\sqrt{\gamma_2} P_{th,2}}{C(\omega)+D(\omega)}, \label{fadd}
\end{eqnarray}
The first two terms describe the input noise from the cavity, which cause the SQL \cite{NJP.10.095010}.
The third term describes the thermal noise from the mechanical environment.
According to the expression in \esref{eqps}, we can find that $A(\omega)$, $B(\omega)$, $C(\omega)$ and $D(\omega)$ contain coherent terms caused by coupling $V$, i.e. $V\chi_{m1}\chi_{m2}$.
Thus, we have the ability to reduce the value of $F_{add}$ by selecting the appropriate coupling  rate $V$.
From the general definition of the noise spectrum, we have
\begin{equation}
S_{add}(\omega )=\frac{1}{2}[S_{FF}(\omega )+S_{FF}(-\omega )],
\end{equation}
where $S_{FF}(\omega )=\int d\omega ^{\prime }\langle F_{add}(\omega)F_{add}(\omega ^{\prime })\rangle $.
The vacuum radiation input noise $a_{in}$ satisfy $\delta $-correlation function.
The operator $P_{th,j}(\omega)$ describes the input thermal noise, under Born-Markov approximation, we have $\langle P^{\dag}_{th,j}(\omega) P_{th,j}(\omega') \rangle\approx n_{th,j}\delta(\omega-\omega')$,
where $n_{th,j}=[\exp(\hbar \omega_{m,j}/k_B T)-1]^{-1}$ describes the equivalent thermal occupation.
The additional noise spectrum density becomes
\begin{eqnarray}
S_{add}(\omega)&=& \frac{1}{2}[|\frac{A(\omega)}{E(\omega)}|^2+|\frac{B(\omega)}{E(\omega)}|^2]+S_{th}(\omega), \label{saddex}
\end{eqnarray}
where $E(\omega)=C(\omega)+D(\omega)$, $S_{th}(\omega)=\gamma_1 n_{th1}|C(\omega)/E(\omega)|^2+\gamma_2 n_{th2}|D(\omega)/E(\omega)|^2$.
Under the condition $\Delta'=\omega_{m,j}=\omega_m$ and $G_j=G$, we have
\begin{widetext}
\begin{eqnarray}
S_{add}(\omega)&=&|\frac{1-V^2\chi_{m1}\chi_{m2}}{2V\chi_{m1}\chi_{m2}-\chi_{m1}-\chi_{m2}}\frac{-i[\omega^2+(\kappa/2-i\omega)^2]}{2\sqrt{\kappa}(\kappa/2-i\omega)G} +\frac{2(\kappa-i\omega_m)}{\sqrt{\kappa}(\kappa-2i\omega_m)}G|^2+S_{th}(\omega).
\end{eqnarray}
\end{widetext}
This should be compared to the result of the standard optomechanical scenario (SOM) \cite{RevModPhys.86.1391}
\begin{eqnarray}
S_{add1}(\omega)&=&|\frac{-i[\omega^2+(\kappa/2-i\omega)^2]}{2\chi_{m1}\sqrt{\kappa}(\kappa/2-i\omega)G} \nonumber\\
&&+\frac{2(\kappa-i\omega_m)G}{\sqrt{\kappa}(\kappa-2i\omega_m)}|^2+\gamma_1 n_{th1},
\end{eqnarray}
with its familiar shot-noise term scaling as $1/G^2$ and the back-action term scaling as $G^2$.
The SQL of SOM ($S_{SQL,1}$) and OMDP ($S_{SQL,2}$) can be obtained by minimizing $S_{add1}$ and $S_{add}$ with respect to $G$ at $T=0$.
The minimal value of $S_{SQL,1}(\omega)$ and $S_{SQL,2}(V=0,\omega)$ can be obtained at $\omega=\omega_m$.
To investigate the noise limit of OMDP, we define two SQL proportion factors,
\begin{eqnarray}
R_1(\omega)&=&\frac{S_{SQL,2}(\omega)}{S_{SQL,1}(\omega=\omega_m)},\\
R_2(\omega)&=&\frac{S_{SQL,2}(\omega)}{S_{SQL,2}(V=0,\omega=\omega_m)},
\end{eqnarray}
$R_1$ is defined to compare the SQL of OMDP and SOM.
$R_1<1$ denotes OMDP can beyond the SQL of SOM.
$R_2$ is defined to compare the SQL of OMDP with and without mechanical interaction.
$R_2<1$ denotes OMDP with mechanical interaction can beyond the SQL of that without mechanical interaction.
The comparison of SQL with different parameters are shown in \fref{fig3}.
The white-dashed line denotes $R_j=1$.
We can see that the additional noise exhibit a minimal value at the specific frequency in the dark area in the figure.
Under the condition $\{G_j,\gamma_{m,j} \}\ll \omega_m$, this optimized frequency can be obtained as $\omega_{eff}\approx \sqrt{\omega_m(\omega_m+V)}$.
It is shown that, the SQL of OMDP can go beyond the SQL of the SOM and OMDP without mechanical interaction.
The corresponding optimized parameters area can be found between the white-dashed lines in the figure.
At the optimized frequency $\omega_{eff}$, $R_j$ can reaches rather low value, where $R_1$ is on the order $10^{-7}$ and $R_2$ is on the order $10^{-5}$.
In \fref{fig3}(a), when mechanical interaction $V=0$, $R_1\lesssim 1$, the SQL
of OMDP is little lower than SOM in frequency area $\omega \in [0.9\omega_m,1.1\omega_m]$.
With the increases of $V$, $R_j$ decreases.
Similar characters can be found of $R_2$ in \fref{fig3}(b), with the help of mechanical interaction the SQL of OMDP can reaches rather low level.

In above calculation, the discussion of SQL is under the condition $T=0$.
For the real situation, the thermal noise should be considered.
It is noteworthy that, in our system, the addition of detection ports does not increase the effective thermal noise, but makes the thermal noise slightly reduced.
According to the expression of $S_{th}(\omega)$ under \eref{saddex}, the thermal noise will become $1/2$ of the original one ($S_{th}(\omega)= \gamma_1 n_{th1}/2$) when the parameters of the oscillators are identical, because there is no quantum correlation between the thermal operator of different oscillators.

\begin{figure}
  \centering
  \includegraphics[width=8cm]{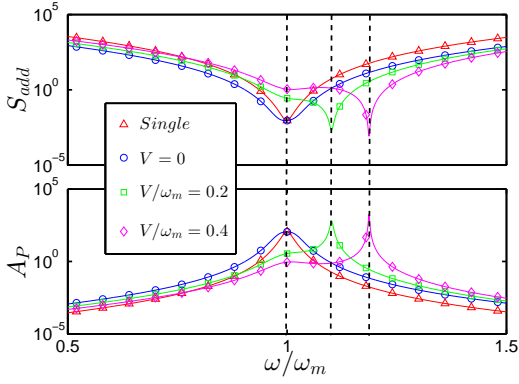}\\
  \caption{Additional noise and amplification coefficient as a function of frequency $\omega$.
 $G/\omega_m=0.03$, $\kappa/\omega_m=0.1$, $n_{th}=10$,  $\Delta \omega/\omega_m=0$, $\gamma_1=\gamma_2=10^{-5}\omega_m$, $\omega_m=5\times10^{6}Hz$. }\label{fig2}
\end{figure}

\begin{figure}
  \centering
  \includegraphics[width=8cm]{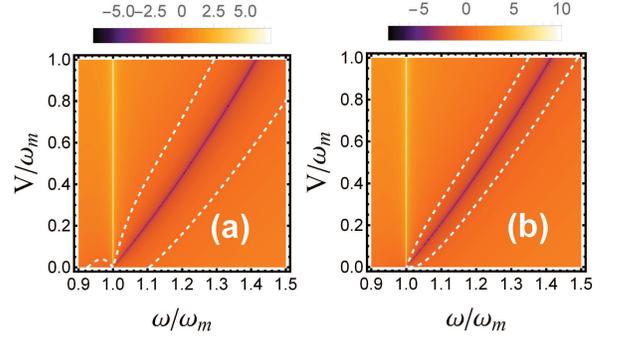}  \\
  \caption{$R_j$ as a function of frequency $\omega$ and mechanical coupling $V$ (on a logarithmic scale).
  White-dashed line denotes $R_j=1$.
  (a) and (b) is labeled as density plot of $R_1$ and  $R_2$, respectively.
  Other parameters are same with \fref{fig2}.}\label{fig3}
\end{figure}

The additional noise spectrum and amplification spectrum are shown in \fref{fig2}.
There is a minimal value at the frequency $\omega_{eff}$ of additional noise.
And we can get lower additional noise with larger value of $V$.
When $V=0$, the value of minimal additional noise is almost the same as that of SOM with single detection port (mark as 'single').
The same characters can be found in amplification spectrum, but the larger the value of $V$, the larger the amplification rate can be get.

\begin{figure}
  \centering
  \includegraphics[width=8cm]{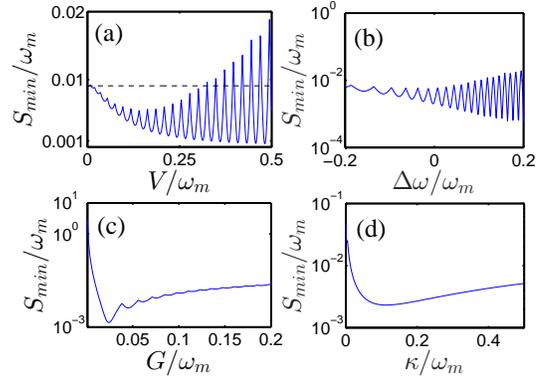}\\
  \caption{(a) The minimal additional noise as a function of $V$, $G/\omega_m=0.03$, $\kappa/\omega_m=0.1$, $\Delta \omega/\omega_m=0$.
  (b)The minimal additional noise as a function of $\Delta \omega$, $G/\omega_m=0.03$, $V/\omega_m=0.2$, $\kappa/\omega_m=0.1$.
  (c) The minimal additional noise as a function of $G$, $V/\omega_m=0.2$, $\kappa/\omega_m=0.1$, $\Delta \omega/\omega_m=0$.
  (d)The minimal additional noise as a function of $\kappa$, $G/\omega_m=0.03$, $V/\omega_m=0.2$, $\Delta \omega/\omega_m=0$,.
   Other parameters are $\gamma_1=\gamma_2=10^{-5}\omega_m$, $\omega_m=5\times10^{6}Hz$, $n_{th}=10$.}\label{fig4}
\end{figure}

To study the the relationship between additional noise and adjustable parameters, we plot \fref{fig4}.
As shown in \fref{fig4}(a), the minimum additional noise $S_{min}$ exhibits an oscillating decrease with the increase of V, and the amplitude of turbulence also goes larger.
This is consistent with the analysis of \eref{fadd}, the coupling between two oscillators can cause interference cancellation of the additional noise.
But the value of $V$ is not the larger the better, negative progress of detection like destructive interference also exist.
For example, when $V=0.47 \omega_m$, the minimum noise $S_{min}=0.015$, which is larger than the case of that without mechanical interaction.
Thus, in order to suppress additional noise, we need to select the appropriate coupling coefficient.
As shown in \fref{fig4}(b), we plot the evolution characteristics of additional noise with frequency difference $\Delta \omega=\omega_{m1}-\omega_{m2}$.
Similar to the conclusion of \fref{fig4}(a), the additional noise exists the coherent effect due to mechanical coupling.
The frequency difference $\Delta \omega$ will affect the value of $\chi_{m2}$ and then affect the coherence term, i.e. $V \chi_{m1}\chi_{m2}$.
The amplitude of the curve increases with the increase of $\Delta \omega$.
Thus, to optimize the additional noise, we need not only select the appropriate coupling but also select the appropriate frequency difference.
In the experimental realization, to obtain larger mechanical coupling, we usually need to set $\Delta \omega\approx 0$ \cite{PhysRevLett.115.017202}.
As shown in \fref{fig4}(c), we investigated the effect of linearized coupling strength on minimum additional noise.
The minimum additional noise $S_{min}$ has a valley value along with the increase of the linear coupling coefficient, and appears at $G=0.02\omega_m$.
When $G<0.02\omega_m$, $S_{min}$ decreases with the increase of $G$. When $G>0.02\omega_m$, $S_{min}$ increases with the increase of $G$.
This phenomenon is due to the existence of a dependent relationship between the feedback noise and the shot noise, which is dependent on the linearized coupling strength, which have been studied in recent researches \cite{NJP.19.083022}.
As shown in \fref{fig4}(d), we plot $S_{min}$ as function of $\kappa$.
Similar to the conclusion of \fref{fig4}(c), due to the competitive relationship between the feedback noise and the shot noise, $S_{min}$ has a minimum value for $\kappa$.

\begin{figure}
  \centering
  \includegraphics[width=7.5cm]{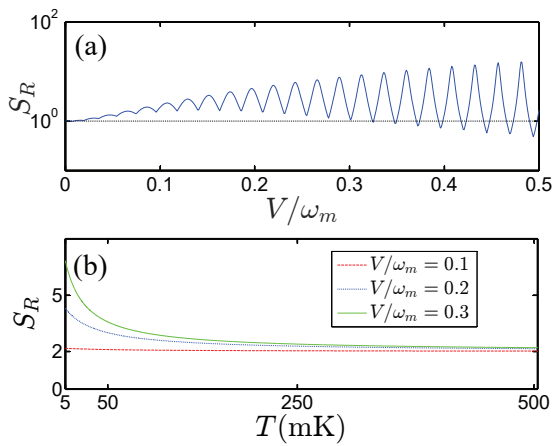}\\
  \caption{(a) $S_R$ as a function of coupling $V$, where environment temperature $T=1 \text{mK}$.
  (b) $S_R$ as a function of environment temperature $T$ with different coupling.
Other parameters are smae with \fref{fig2}}\label{fig5}
\end{figure}

A quantum detector based on optomechanical system should not be benchmarked by its additional noise suppression alone, but instead one should measure its performance by considering also signal-to-noise ratio (SNR), and the linearity of the detector response.
According to the standard SNR definition and the expression of \eref{mout1}, we have $SNR(\omega)=\xi^2 B^2/S_{add}(\omega)$.
Where $\xi$ and $B$ are constant in our analysis.
Thus, we can easily learn the characteristics of SNR from the characteristics of additional noise.
To investigate the enhancement of SNR in our scheme under finite temperature, we define the proportion factor $S_R=SNR(\omega_{eff})/SNR_0$, where $SNR_0$ is the maximum SNR of SOM.
When $S_R>1$ indicates that when the coupling is present, the SNR is better than the condition without mechanical interaction, and vice versa.
As shown in \fref{fig5}(a), at low temperature regime, our scheme exhibit remarkable advantage with appropriate coupling rate, which is consistent with the analysis in \fref{fig4}(a).
As shown in \fref{fig5}(b),
with the increase of temperature, $S_R$ gradually decreases and tends to a constant $2$.
Larger coupling could have larger value of $S_R$ and stronger ability to resist environment noise.
When the environment temperature is high enough, the value of $S_R$ depends only on the thermal noise of the mechanical oscillators.
In the previous analysis, we know that our system can reduce the effective thermal noise by half, so the final $S_R$ will approach to the constant $2$.
Thus the design of mechanical oscillators on the coupling is the key to use mechanical correlation to enhance SNR and reduce additional noise.
Compared with SOM, although our scheme has a greater advantage in resisting noise, it is still necessary to keep the detector in a low temperature environment.

\section{Application example: weak magnetic field detection}
Taking magnetic field as an example, our system can be used as a magnetometer.
The magnetic field can be responsive to the mechanical oscillators with surface charge \cite{NM.11.30}.
As shown in \fref{fig1}, oscillator 1 and 2 are simultaneously used as probes for magnetic field.
Different strength of the magnetic field will cause different displacement of the oscillators.
Assuming that the magnetic field is homogeneous, the corresponding Hamiltonian can be written as $H_b = -\sum_{j=1,2}B\xi_j q_j$, where $B$ represents the intensity of the magnetic field, $\xi_j=I L$ denotes the constant related to the surface charge characteristics, where $I$ denotes surface current of the mechanical oscillators and $L$ denotes the size of the oscillators.

In macro-optomechanical system, we have mechanical frequency $\omega_m=2 \pi \times 10.56 \text{MHz}$, $\gamma_m=2\pi \times 32 \text{Hz}$, mechanical diameter $15\mu \text{m}$ \cite{nature.475.359}.
\begin{figure}
  \centering
  \includegraphics[width=8cm]{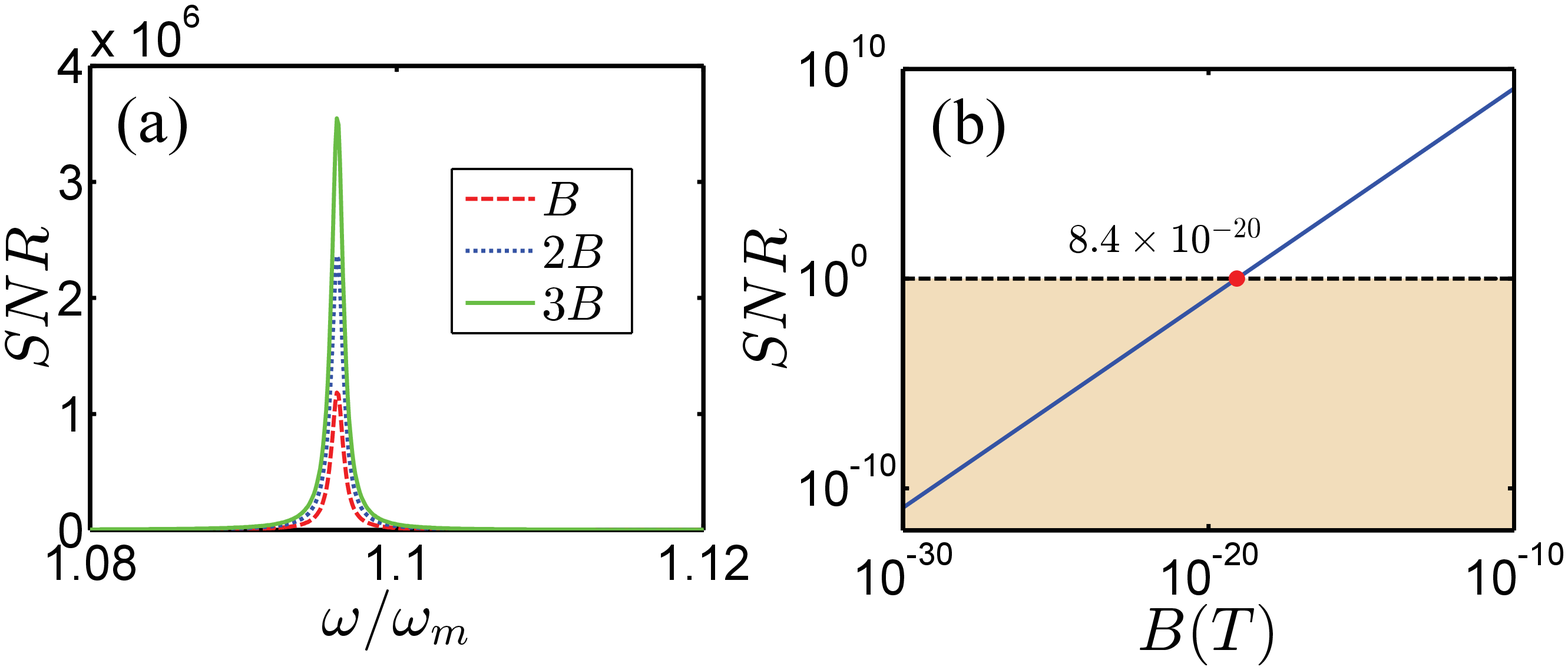}\\
  \caption{(a) SNR spectrum with different magnetic field intensity.
  (b) $SNR(\omega_{eff})$ as a function of magnetic field intensity.
   $B=10^{-13}\text{T}$, $I=10\mu\text{A}$, $V/\omega_m=0.2$, $\kappa/\omega_m =0.1$, $\gamma_1=\gamma_2=2\pi\times32\text{Hz}$, $\omega_m=2\pi\times10.56\text{MHz}$, $T=1\text{mK}$.}\label{fig6}
\end{figure}
The SNR and precision of the detection system has been shown in \fref{fig6}.
Consistent with the conclusion of the previous section, the optimal detection frequency is $\omega_{eff}$.
The SNR of the field detector can reach the order $10^{6}$, when $B=10^{-13}\text{T}$, and the corresponding output photon number is $n=1.7\times 10^{6}$.
With the increase of field strength, the SNR is increases significantly.
In \fref{fig6}(b), we examined the detection accuracy and response characteristics of the detector.
Here, we define the detection accuracy is the intensity of magnetic field when SNR equal to 1.
It can be found that, under a given parameters, the detection accuracy can reach $8.4\times 10^{-20}\text{T}$.
Besides, there is an obvious linear relationship between SNR and input field strength $B$.
In our calculation, the additional noise does not change with different field strength.
Therefore, SNR can be used directly to indicate the intensity of output signal.
That is to say, our weak field detection system has a good linear response, which is the basic requirement of the detector.
In addition, the SNR can be further increased by increasing the size of the oscillators or the current through therm.
\section{conclusion }
In conclusion, the weak-field dual-port detection protocol can effectively suppress the additional noise and even break through the standard quantum limit of the standard optomechanical protocol \cite{RevModPhys.86.1391}.
By using the quantum correlation between mechanical oscillators and selecting the appropriate coupling strength, we can greatly reduce the noise without weakening the signal and make the system achieve a high SNR.
Under the existing experimental conditions, we take the weak magnetic field as an example to investigate the performance of our scheme.
Under the given parameters in the Ref. \cite{nature.475.359}, the simulated detection accuracy can reach $8.4\times 10^{-20}\text{T}$.
The detector also has a high SNR and a good linear response curve.
Our scheme provide a promising application of the optomechanical system in quantum weak field detection.

\section{Acknowledgments}
Project supported by the National Natural Science Foundation of China (Grant Nos. 11704026, 11304174, 11704205, U1530401), the China Postdoctoral Science Foundation funded project  (Grant No. 2018T110039), the Natural Science Foundation of Shandong Province (Grant No. ZR2013AQ010) and the Natural Science Foundation of Ningbo (Grant No. 2018A610199).
\bibliography{FDE}
\end{document}